\documentclass[aps,prd,twocolumn,showpacs,preprintnumbers,amsmath,amssymb,longbibliography]{revtex4-1}
\usepackage{hyperref}
\usepackage{bm,bbm}
\usepackage{graphics}
\usepackage{braket}
\usepackage{mathtools}
\usepackage{mathrsfs}
\usepackage{comment}
\usepackage{autobreak}
\usepackage{color}

\edef\restoreparindent{\parindent=\the\parindent\relax}
\usepackage{parskip}
\restoreparindent

\usepackage{tikz}
\usetikzlibrary{arrows,arrows.meta,intersections, calc,positioning,decorations.pathreplacing,decorations.pathmorphing,shapes}
\usetikzlibrary{patterns}
\usetikzlibrary{decorations.markings}
\usetikzlibrary{knots}

\begin{document}

\title{Complex saddles of three-dimensional de Sitter gravity via holography}
\preprint{YITP-23-16}

\author{Heng-Yu Chen,$^{a}$ Yasuaki Hikida,$^b$ Yusuke Taki$^b$ and Takahiro Uetoko$^c$}

\affiliation{$^a$Department of Physics, National Taiwan University, Taipei 10617, Taiwan}
\affiliation{$^b$Center for Gravitational Physics and Quantum Information, Yukawa Institute for Theoretical Physics, Kyoto University, Kyoto 606-8502, Japan}
\affiliation{$^c$Science in General Education, National Institute of Technology, Kushiro College, Hokkaido 084-0916, Japan}


\begin{abstract}

We determine complex saddles of three-dimensional gravity with a positive cosmological constant by applying the recently proposed holography. It is sometimes useful to consider a complexified metric to study quantum gravity as in the case of the no-boundary proposal by Hartle and Hawking. However, there would be too many saddles for complexified gravity, and we should determine which saddles to take. We describe the gravity theory by three-dimensional SL$(2,\mathbb{C})$ Chern-Simons theory. At the leading order in the Newton constant, its holographic dual is given by Liouville theory with a large imaginary central charge. We examine geometry with a conical defect, called a de Sitter black hole, from a Liouville two-point function. We also consider geometry with two conical defects, whose saddles are determined by the monodromy matrix of Liouville four-point function. Utilizing Chern-Simons description, we extend the similar analysis to the case with higher-spin gravity.

\end{abstract}

\maketitle

\section{Introduction}

When considering the path integral for quantum gravity, it is sometimes useful to analytically continue the geometries to complex ones; however, as recently pointed out in \cite{Witten:2021nzp}, not all complexifications of a real geometry yield physically sensible answer.
A nice canonical example is provided by the geometry obtained from complexifying $(d+1)$-dimensional sphere $(S^{d+1})$ metric $ds^2 = \ell^2 ( \theta  '(u) ^2 d u ^2 + \cos ^2 \theta  (u) d \Omega ^2_d)$. Here $\ell$ is a real length scale and $d\Omega ^2_d$ is a metric of $S^d$, while $\theta$ and $u$ are both complex valued such that $\theta(u)$ gives the immersion of the resultant metric into the complex manifold. We now consider the no-boundary proposal by Hartle and Hawking \cite{Hartle:1983ai}, where the universe starts from nothing. For complex $\theta(u)$, the universe can start from any of $\theta = (n + 1/2) \pi$ with $n \in \mathbb{Z}$ and then go to $i \infty$, this may imply that we need to sum over infinite number of saddle points when evaluating the path integral, which are obviously too many.
However using the criterion that all exact forms have norms with non-negative real parts as proposed in \cite{Louko:1995jw,Kontsevich:2021dmb,Witten:2021nzp}, 
it was shown recently in \cite{Witten:2021nzp} that the allowable geometry here is only given by $n = -1,0$ saddle points, which are exactly the geometry analyzed in \cite{Hartle:1983ai}.

The non-negative real norm criterion can be applied to many examples of complex analytic continuation in principle, but it can be difficult to implement explicitly.
In this letter, we propose a different approach to identify the appropriate saddle points of complex geometry path integral for quantum gravity theories from their holographic duals.
Explicitly we consider a simple setup, i.e. three-dimensional gravity with positive cosmological constant and derive its complex saddle points. The gravity theory can be described by SL$(2,\mathbb{C})$ Chern-Simons gauge theory \cite{Witten:1988hc,Witten:1989ip,Witten:2010cx}. Recently, it was proposed that the leading effects in the Newton constant $G_N$ can be captured by a specific limit of  holographic dual field theory, i.e. Liouville theory \cite{Hikida:2021ese,Hikida:2022ltr,Chen:2022ozy,Chen:2022xse}.
See, e.g. \cite{Basile:2023ycy} for a related work on three-dimensional gravity with negative cosmological constant.

The holography considered here is in fact an explicit example of the general proposal posed abstractly in \cite{Maldacena:2002vr}, see also \cite{Strominger:2001pn,Witten:2001kn}.
We prepare the Hartle-Harking wave functional of the universe as $\Psi[\chi^{(0)}_j] = \int \mathcal{D} \chi_j \exp (i S_\text{dS}[\chi_j])$. Here $S_\text{dS}[\chi_j]$ is the action of gravity theory on de Sitter (dS) space-time and fields $\chi_j$ of the theory are required to satisfy the boundary conditions $\chi_j = \chi_j^{(0)}$ at future infinity. We consider the geometry created due to the back reaction of a scalar field with large energy $E$. Suppose that the saddle points of the path integral are realized by $\chi_j = \chi_j^{n}$ with label $n$. Then the wave functional can be written as 
\begin{align}
\Psi \sim \sum_n \exp \left( S_\text{GH}^{(n)}/2 + i \mathcal{I}^{(n)} \right) \label{Psi}
\end{align}
at the semi-classical limit. The fields are assumed to take complex values, which leads to the complex action as in \eqref{Psi} with real $S_\text{GH}^{(n)}$ and $\mathcal{I}^{(n)}$. The largest $S_\text{GH}^{(n)}$ gives the dominant contribution to  Gibbons-Hawking entropy associated with the geometry \cite{Bekenstein:1973ur,Gibbons:1976ue,Gibbons:1977mu}.

According to the proposal of \cite{Maldacena:2002vr}, the wave functional of the universe can be evaluated with the correlation functions of dual conformal field theory (CFT). As mentioned above, we introduce a bulk scalar field with energy $E$, which creates a back reacted geometry. The dual CFT operator $V_h$ should have conformal weight $h = i h^{(g)}$ with real $h^{(g)}$ with $2 h^{(g)} = \ell E$. The central charge of dual CFT $c = i c^{(g)}$ $(c^{(g)} \in \mathbb{R})$ is related to gravity parameters as $c^{(g)} = 3 \ell/(2 G_N)$ \cite{Strominger:2001pn}.
Since the bulk scalar field connects two boundary points, the configuration should be related to two-point function of the dual CFT operators as 
\begin{align}
\begin{aligned}
\Psi &= \langle V_h (z_1) V_h (z_2) \rangle  \\
     &= \int \mathcal{D} \phi_j e^{- S_\text{CFT} [\phi_j]} V_h (z_1) V_h (z_2)\, .
\end{aligned}
\end{align}
Here we denote the dual CFT fields and their action by $\phi_j$ and $S_\text{CFT}[\phi_j]$, respectively. We assume that the conformal weight satisfies $h \sim c$, and in that case the insertion of vertex operators can be regarded as a part of modified action. Suppose that the saddle points of the modified action are given by $\phi_j = \phi_j^{n}$. Then the two-point function can be put into the form of \eqref{Psi}, from which we can read off the map between the saddle points of gravity theory and dual CFT. In general, quantum gravity is not well-defined, at least non-perturbatively, but its dual CFT is well-formulated and can be analyzed more deeply. Therefore, our holographic approach provides useful insights on allowed geometry. Above, we explained the case with CFT two-point functions for simplicity, but the procedure can be extended to the case with CFT multi-point function as studied below.

Specifically, we consider three-dimensional de Sitter (dS$_3$) space-time with a conical defect, which is often called as dS$_3$ black hole \cite{DESER1984405} in this letter. There are additional complex saddles of Chern-Simons theory related by large gauge transformations. We determine which saddles to take from the semi-classical analysis of the two-point function in Liouville theory by  \cite{Harlow:2011ny}.
We also deal with geometry including two conical defects constructed in \cite{Hikida:2021ese,Hikida:2022ltr}, where the dual CFT partition functions were obtained in terms of modular $S$-matrix \cite{Witten:1988hf}. Here we derive the same relation from the monodromy matrix of four-point functions of Liouville theory.
Utilizing the Chern-Simons description, we extend the analysis to the case with higher-spin gravity as well.

\section{Three-dimensional dS black hole}

The metric of black hole solution on dS$_3$ is given by
\begin{align}
\begin{aligned}
&ds^2= \ell^2 \left[ dr^2 /(1 - 8 G_N E - r^2) \right . \\ & \qquad \qquad \left .
- (1 - 8 G_N E - r^2) dt^2 + r^2 d \phi ^2 \right] \, .
\end{aligned}\label{dSBHmetric}
\end{align}
Here $\ell$ is the radius of dS$_3$ and $E$ is the energy of an excitation \cite{Spradlin:2001pw}.
The periodicity is assigned as $\phi \sim \phi + 2 \pi$ and the horizon is located at $r = \sqrt{1 - 8 G_N E}$.
We may consider a Wick rotation as $i t = t_E$, then the smoothness at the horizon requires the periodicity
$t_E \sim t_E + 2 \pi /\sqrt{1 - 8 G_N E}$.
The Gibbons-Hawking entropy associated with the horizon is \cite{Bekenstein:1973ur,Hawking:1975vcx,Gibbons:1977mu,Gibbons:1976ue}
\begin{align}
S_\text{GH}  =  \frac{ \pi \ell \sqrt{1 - 8 G_N E} }{ 2 G_N}\, . \label{BHentropy}
\end{align}

We describe the gravity theory by SL$(2,\mathbb{C})$ Chern-Simons gauge theory with the action \cite{Witten:1988hc}
\begin{align}
\begin{aligned}
&S = S_\text{CS} [A] - S_\text{CS} [\tilde A] \, , \\
&S_\text{CS}[A]  = - \frac{\kappa}{4 \pi} \int \text{tr} \left( A \wedge d A + \frac{2}{3} A \wedge A \wedge A \right) \, .
\end{aligned}
 \label{CSaction}
\end{align}
The Chern-Simons level $ \kappa \, (\in \mathbb{R}) $ is related to the gravity parameters as $\kappa = \ell /(4 G_N)$.
As explained in \cite{Witten:1989ip,Witten:2010cx}, we treat $A , \tilde A$ as two independent one-form fields taking values in $\mathfrak{sl}(2,\mathbb{C})$. For convention, we use the generators of $\mathfrak{sl}(2)$ Lie algebra given by $L_0,L_{\pm 1}$ satisfying $[L_n , L_m] = (n - m)L_{n+m}$. We choose its normalization as $\text{tr} (L_0  L_0) = 1/2$ and their complex conjugations as $(L_0)^* = - L_0$, $(L_{\pm1})^* =  L_{\mp 1}$. If we choose a real slice $\tilde A = - A^*$, then the equation of motion is the same as the Einstein equation with positive cosmological constant for Lorentzian space-time.

The solutions to the equations of motion are given by flat connections.
We put the gauge fields in  the form
\begin{align}
\begin{aligned}
&A = e^{- i \theta  L_0} a  e^{ i \theta L_0} + i L_0 d \theta \, , \\
&\tilde A = - e^{i \theta L_0} \tilde a  e^{- i \theta L_0} - i L_0 d \theta \, .
\end{aligned}
\label{gaugedS}
\end{align}
We consider  a solution
\begin{align}
\begin{aligned}
&a  = i \sqrt{\frac{ 2 \pi \mathcal{L}}{\kappa}} \left( L_1 -    L_{-1} \right) (d \phi + id t ) \, ,  \\
&\tilde a  =  -  i \sqrt{\frac{ 2 \pi \mathcal{ L}}{\kappa}} \left( L_{-1} - L_{1} \right) (d \phi - id t ) \, . 
\end{aligned}
\label{dSBHgauge}
\end{align}
The bulk metric can be read off from
\begin{align}
g_{\mu \nu} = - \frac{\ell^2}{2} \text{tr} (A_\mu - \tilde A_\mu) (A_\nu - \tilde A_\nu) \, , \label{metricdS}
\end{align}
which leads to
\begin{align}
ds^2 = \ell^2 \left[ d \theta^2 - \frac{8 \pi \mathcal{L} }{\kappa} \sin ^2 \theta d t^2 
+ \frac{8 \pi \mathcal{L}}{\kappa} \cos ^2 \theta d\phi^2  \right] \, .
\end{align}
The coordinate transformation $r = \sqrt{8 \pi \mathcal{L}/\kappa}\cos \theta$ with $8 \pi \mathcal{L}/\kappa = 1 - 8 G_N E$ maps the above metric to the one in \eqref{dSBHmetric}.
Performing the Wick rotation as $i t \to t_E$, the smoothness at the horizon requires the periodicity 
$t_E \sim t_E + \sqrt{ \pi \kappa /2\mathcal{L}} $ as before.
Note that the Wick rotation breaks the condition $\tilde A = - A^*$, and the corresponding geometry is now complexified.

In order to characterize the black hole geometry in terms of Chern-Simons theory, it is convenient to introduce a holonomy matrix along the time-cycle as \cite{Gutperle:2011kf,Ammon:2012wc}
\begin{align}
\mathcal{P} e^{\oint A} = \mathcal{P} e^{ \oint d t_E A_{t_E}}  = e^{- i \theta   L_0} e^\Omega e^{i \theta  L_0} \,  . \label{dShol}
\end{align}
Here $\mathcal{P}$ indicates the path ordering. The eigenvalues of $\Omega$ for the configuration \eqref{dSBHgauge} are $( \pi i , - \pi i)$.
As saddle points, we pick up non-singular geometry in the sense that the holonomy matrix is trivial, i.e., $\pm \mathbbm{1}$.
It can be realized also by the cases with eigenvalues $(2\pi \tilde n i, - 2 \pi \tilde n i)$, where $\tilde n \in \mathbb{Z}$ or $\tilde n \in \mathbb{Z} + 1/2$. 
A configuration of gauge fields with the holonomy matrix may be given by 
\begin{align}
\begin{aligned}
&a  = i \sqrt{\frac{2 \pi \mathcal{L}}{\kappa}} \left( L_1 -    L_{-1} \right) ( d \phi + 2 \tilde n dt_E ) \, ,  \\
&\tilde a  =  -  i \sqrt{\frac{2 \pi \mathcal{\tilde L}}{\kappa}} \left( L_{-1} - L_{1} \right) ( d \phi - 2 \tilde n d t_E ) \, . 
\end{aligned}
\end{align}
The metric from the configuration can be read off as
\begin{align}
ds^2 = \ell^2 \left[ d \theta^2 + \frac{8 \pi {(2{\tilde n})^2} {\mathcal L}}{\kappa} \sin ^2  \theta  d t^2_E  +  \frac{8 \pi {\mathcal L}}{\kappa} \cos ^2 \theta  d\phi^2 \right] \, .
\end{align}
The complex geometry with the metric contributes to the real part of \eqref{Psi} and its Gibbons-Hawking entropy  is evaluated as
\begin{align}
 S_\text{GH}^{(\tilde n)} = 8 \pi \tilde n \sqrt{2 \pi \kappa \mathcal{L}} = 2 \tilde n\frac{ \pi \ell \sqrt{1 - 8 G_N E}}{2 G_N} \, . \label{extsol}
\end{align}
See \cite{Hikida:2021ese,Hikida:2022ltr} for the case with conical defect geometry.

In this way, the saddle points of the Chern-Simons theory can be labeled by $\tilde n$. Solutions with different $\tilde n$ are related by large gauge transformation as the holonomy condition suggested. As pointed out in \cite{Witten:2010cx,Harlow:2011ny}, the large gauge transformation is not a symmetry of the complexified theory but generates new saddles. As mentioned above, the saddle points associated with $\tilde n = \pm 1/2$ are allowed geometries of \cite{Witten:2021nzp}. These two should have the same geometrical interpretation, since they can be mapped by replacing $t_E$ with $- t_E$. 
In the following, we obtain the same conclusion via holography.

\section{Dual Liouville field description}

The action of Liouville theory is given by
\begin{align}
S_\text{L} = \frac{1}{2 \pi} \int d^2 z \sqrt{\tilde g} \left[ \partial \phi \bar \partial \phi  + \frac{Q}{4} \tilde{\mathcal{R}} \phi +  \pi \mu e^{2 b \phi}\right] \, .
\end{align}
The ``physical'' metric is given as $g_{ij} = e^{\frac{2}{Q} \phi} \tilde g_{ij}$.
We mainly work with the flat reference metric such that the curvature is $\tilde{\cal R} = 0$.
The vertex operators are defined by
$
V_\alpha = e^{2 \alpha \phi}
$
with conformal weights $h = \bar h = \alpha (Q - \alpha)$. The central charge $c$ is related to the background charge 
$Q = b + b^{-1}$ as
$
c= 1 + 6 Q^2  \, .
$
In order to obtain finite action, we also need to add proper boundary terms and assign boundary conditions as explained in \cite{Zamolodchikov:1995aa,Harlow:2011ny}.

We are interested in the regime with a $c^{(g)} (\equiv - i c)$ that is real and very large, thus we may approximate as follows (see \cite{Hikida:2022ltr}):
\begin{align}
b^{-2} =  \frac{i c^{(g)}}{6}   - \frac{13}{6}  + \mathcal{O} ((c^{(g)})^{-1})\, ,
\end{align}
which indicates $b \sim 0$. The  contribution of order $\mathcal{O}((c^{(g)})^0)$ implies that $\text{Re} \,  b^{-2} < 0$.
For $b \sim 0$ with $\phi_c = 2 b \phi$, the action may be written as
\begin{align}
b^2 S_\text{L} &= \frac{1}{8 \pi} \int d^2 z [\partial \phi_c \bar \partial \phi_c + 4 \lambda e^{\phi_c}] \, .
\end{align}
For our purpose, it is convenient to choose $\lambda \equiv \pi \mu b^2$ real and finite, see \cite{long} for the details.

We evaluate the two-point function of heavy operators,
\begin{align}
\begin{aligned}
&\left \langle V_{\alpha} (z_1) V_\alpha (z_2) \right \rangle 
\\& \quad \equiv 
\int \mathcal{D} \phi_c e^{- S_\text{L}} \exp \left( b^{-1} \alpha (\phi_c (z_1) + \phi_c (z_2))\right) \, .
\end{aligned}
\end{align}
A heavy operator is defined with $\alpha = \eta /b$, where we choose $0 \leq \eta \leq 1/2$ for $b \sim 0$. The parameter $\eta$ is related to $E$ in \eqref{dSBHmetric} as $1 - 2 \eta =\sqrt{1 - 8 G_N E}$, see, e.g., \cite{Hikida:2021ese,Hikida:2022ltr}.
We may regard the insertions of heavy operators as a part of action.
Then, the equation of motion becomes
\begin{align}
\partial \bar \partial \phi_c = 2 \lambda e^{\phi_c} -2 \pi \eta [\delta^{(2)} (z- z_1) + \delta^{(2)} (z - z_2) ] \, .
\end{align}
Notice that the equation is invariant under the constant shifts $\phi_c \to \phi_c + 2 \pi i n$ with integer $n$. Therefore, once $\phi_c^0$ is a 
solution to the equation of motion, then the same is true for
$
\phi_c^n = \phi_c^0 + 2\pi i n
$.
The classical action with $\phi_c^n$ was evaluated in \cite{Harlow:2011ny} as
\begin{align}
&b^2 S_\text{L} = 2 \pi i (n+ 1/2)  (1- 2 \eta) + (2 \eta -1) \ln \lambda \\
& + 4 (\eta - \eta^2) \ln |z_{12}|  + 2 [ (1 - 2 \eta) \ln (1 - 2 \eta) - (1 - 2 \eta)] \, .  \nonumber 
\end{align}
We thus read off relevant saddles from the exact expression
\begin{align} 
&\langle V_\alpha (z_1) V_\alpha (z_2) \rangle  = |z_{12}|^{- 4 \alpha (Q - \alpha) } \frac{2 \pi}{b^2} [\pi \mu \gamma (b^2)] ^{(Q - 2 \alpha)/b}  \nonumber \\ 
& \quad \times \gamma (2 \alpha /b -1 - 1/b^2)
\gamma(2b\alpha -b^2 ) \delta (0)
\end{align}
by taking the semi-classical limit.
Here we set $\gamma(x) = \Gamma(x)/\Gamma(1-x)$. The delta function comes from $\langle V_\alpha V_{\alpha '} \rangle \propto \delta (\alpha - \alpha ')$.
For $b \sim 0$ with $\text{Re} \,  b^{-2} < 0$, we find  \cite{Harlow:2011ny}
\begin{align}
&\langle V_\alpha (z_1) V_\alpha (z_2) \rangle \sim  |z_{12}|^{- 4 \eta (1 - \eta)/b^2 } \lambda^{(1 - 2 \eta) /b^2}\nonumber\\ & \quad \times
\left( e^{ -  \pi i (1 - 2 \eta) /b^2} - e^{ \pi i (1 - 2 \eta) /b^2} \right) \label{eq:Liouville 2pt} \\ & \quad \times \exp \left \{ - \frac{2}{b^2} \left[( 1 - 2 \eta )\ln (1 - 2 \eta) - (1 - 2 \eta )  \right] \right \} \delta (0) \, . \nonumber
\end{align}
This can be reproduced from the sum of $e^{- S_\text{L}}$ at the saddle points with $n=-1 , 0$, such that the leading $n=-1$ contribution yields the correct Gibbons-Hawking entropy \eqref{BHentropy}.
The answer seems natural since the conformal weight is invariant under the exchange of $\pm (1 - 2\eta)$. However, this is not the case as the usual choice $\text{Re} \, b^{-2} > 0$ forces $n$ to take whole non-negative or non-positive integers \cite{Harlow:2011ny}.
The label $\tilde{n}$ for saddles of Chern-Simons gravity is related via $2 \tilde n = n + 1/2$ or $2 \tilde n = n $. In the current case, the saddle points are given by $\tilde n = \pm 1/2$, which reproduces the allowable geometry mentioned above. Following the analysis in section 6 of \cite{Harlow:2011ny}, even the classical configurations of Liouville field theory could be mapped to those of Chern-Simons theory in this specific example.

Before moving to more complicated examples, we would like to clarify the holography considered here.
In \cite{Hikida:2021ese,Hikida:2022ltr,Chen:2022ozy,Chen:2022xse}, a dS$_3$/CFT$_2$ correspondence was proposed, which can be obtained as an analytic continuation of AdS$_3$ counterpart by \cite{Gaberdiel:2010pz}. In particular, the dual CFT is given by an analytic continuation of Virasoro minimal model, whose correlation functions can be computed by Liouville theory \cite{Creutzig:2021ykz}. The states of the dual CFT belong to degenerate representations in terms of Liouville theory, which are dual to composite particles and/or conical geometries, see, e.g., \cite{Castro:2011iw,Gaberdiel:2012ku,Perlmutter:2012ds}. In the current case, the CFT is deformed by insertions of heavy operators, which are dual to the same gravity theory but on an asymptotic dS$_3$ black hole geometry.

\section{Geometry dual to four-point function}

We consider the complex saddles of Chern-Simons gravity dual to multi-point functions of Liouville theory next. The geometry may be created due to the back reaction of scalar field departing from multi-points of the future boundary and connecting at some bulk points. For instance, the analytic structure of Liouville three-point functions were examined in \cite{Harlow:2011ny}, and results analogous to the case of two-point functions can be obtained \cite{long}. Here we instead focus on complex saddles dual to CFT four-point functions and develop a new method for identifying Chern-Simons gravity solutions corresponding to the insertions of two linked (unlinked) Wilson loops in Euclidean dS$_3$ analyzed in \cite{Hikida:2021ese,Hikida:2022ltr}.

Let us assume that the dual CFT is rational, such as the SU$(N)_k$ Wess-Zumino-Witten model as in \cite{Hikida:2021ese,Hikida:2022ltr} for the time being. 
We define a correlation function as
\begin{align}
C_{ij} (z , \bar z ) = \frac{\left \langle \mathcal{O}^\dagger_i (\infty) \mathcal{O}^\dagger_j (1) \mathcal{O}_i (z) \mathcal{O}_j (0) \right \rangle}{\left \langle \mathcal{O}^\dagger_i\mathcal{O}_i \right \rangle \left \langle \mathcal{O}^\dagger_j\mathcal{O}_j \right \rangle} \, , \label{Cij}
\end{align}
which can be expanded by conformal blocks as
\begin{align}
C_{ij} (z , \bar z ) = \sum_{p} \mathcal{F}^{ii}_{jj} (p|z) \bar{\mathcal{F}}^{ii}_{jj} (p|\bar z) \, .
\end{align}
Here $p$ labels the exchange primary operators with scaling dimensions $(h_p, \bar{h}_p)$. 
For simplicity, let us consider $z \sim 0$, then the function approximates as $C_{ij} (z , \bar z ) \sim 1$.
\begin{figure}
  \centering
  \includegraphics[width=8.5cm]{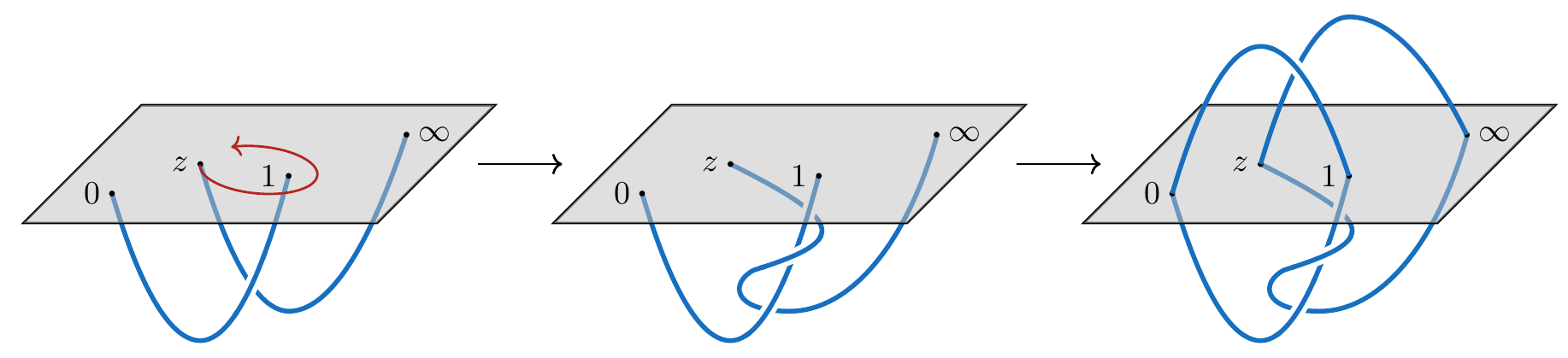}
 \caption{In CFT four-point function, we move $z$ from 0 to 1 and then back to 0 as indicated in the left figure. In terms of Chern-Simons theory, two Wilson lines are winded. Gluing the anti-holomorphic part, two linked Wilson loops are constructed.}
  \label{fig:monodromy}
\end{figure}
As in \cite{Roberts:2014ifa} (see Fig.\,\ref{fig:monodromy}), we start from $z \sim 0$, go around $z =1$ anti-clockwise, then back to $z \sim 0$. 
This move yields a non-trivial monodromy matrix ${\mathcal{M}}_{pq}$ acting on the conformal block as
\begin{align}
\mathcal{F}^{ii}_{jj} (p|z) \to \sum_q \mathcal{M}_{pq} \mathcal{F}^{ii}_{jj} (q|z) \, .
\end{align}
We perform the move only for the holomorphic part and keep the anti-holomorphic part untouched. 
Taking a large central charge limit as in \cite{Hikida:2021ese,Hikida:2022ltr}, such that the scaling dimensions all external operators $h_{i, j}, h_p$ also scale as central charge, 
then the identity block with $h_p=0$ dominates \cite{Hartman:2013mia}. 
Gluing the two parts, we have
\begin{align}
C_{ij} (z , \bar z) \sim \mathcal{M}_{00} \mathcal{F}^{ii}_{jj} (0|z) \bar{\mathcal{F}}^{ii}_{jj} (0|\bar z)
\end{align}
for $z \sim 0$. The monodromy matrix is known to be \cite{Moore:1988uz,Moore:1988ss} (see also \cite{Caputa:2016tgt})
\begin{align}
\mathcal{M}_{00} = \frac{S^*_{ij} S_{00} S_{00}}{S_{00} S_{0i} S_{0j}} \, ,
\end{align}
where $S_{ij}$ is the modular $S$-matrix of CFT character.
As in Fig.\,\ref{fig:monodromy}, the correlator can be interpreted as a partition function of SU$(N)$ Chern-Simons theory with two linked Wilson line loops on $S^3$.  We thus deduce that
\begin{align} \label{S0j}
\left |\left \langle \mathcal{O}^\dagger_j\mathcal{O}_j \right \rangle \right | \sim |S_{0j}| 
\end{align}
and
\begin{align}
\left |\left \langle \mathcal{O}^\dagger_i (\infty) \mathcal{O}^\dagger_j (1) \mathcal{O}_i (z) \mathcal{O}_i (0) \right \rangle \right| \sim |S_{ij}| \, .  \label{Sij}
\end{align}
Here and in the following, we change the normalization of correlators by $|S_{00}|^{-1}$.
These results reproduce those in \cite{Hikida:2021ese,Hikida:2022ltr}.

Let us first comment on the two-point functions.
In the above, we have assumed that CFT is rational.
We may apply the modular $S$-matrix element of Liouville theory for the identity operator and non-degenerate operator \cite{Zamolodchikov:2001ah},
\begin{align}
S_{0j} = -2\sqrt{2} \sin 2 \pi b (\alpha_j - Q/2) \sin 2\pi (\alpha_j - Q/2)/b \, , \label{SijLiou}
\end{align}
then we find
\begin{align} \label{Liouville2pt}
|S_{0j}| \sim \left   | e^{\frac{\pi}{6} c^{(g)} \sqrt{1 - 8 G_N E_j} } -  e^{- \frac{\pi}{6} c^{(g)} \sqrt{1 - 8 G_N E_j} } \right | \, .
\end{align}
The expression reproduces the result obtained by Liouville theory \eqref{eq:Liouville 2pt} including the sub-leading saddle.

We next consider geometry corresponding to two unlinked Wilson loops on $S^3$ in the Chern-Simons theory. We do not perform any move in this case, thus we should have 
$
C_{ij}(z,\bar z) \sim 1
$.
Using \eqref{S0j}, we find
\begin{align}
\left |\left \langle \mathcal{O}^\dagger_i (\infty) \mathcal{O}^\dagger_j (1) \mathcal{O}_i (z) \mathcal{O}_j (0) \right \rangle \right| \sim \left|\frac{S_{0i} S_{0j}}{S_{00}}\right|\, .
\end{align}
This also reproduces a finding in \cite{Hikida:2021ese,Hikida:2022ltr} for two unlinked Wilson loops.

One may be concerned with the assumption of the rationality of dual CFT.
We thus want to reexamine the four-point conformal block $\mathcal{F}^{ii}_{jj} (p|z)$ in terms of Liouville theory. Here we set $\mathcal{O}_a = e^{2 \eta_a \phi /b}$ $(a=i,j)$ with $b \sim 0$.
According to eq.\,(2.43) of \cite{Fitzpatrick:2016mjq}, the conformal block behaves near $z \sim 1$ as
\begin{align} \label{cbsaddle}
\begin{aligned}
 & \mathcal{F}^{ii}_{jj} (p|z) \sim \sum_{m,\delta_\kappa = \pm}c_m^{\delta_\kappa}
 (1-z)^{\frac{c}{6} \kappa_m^{\delta_\kappa }}\, , \\
 &\kappa_m^{\delta_\kappa} = m (1 - m) - \frac12 - \delta_\kappa \frac{(1 - 2 \eta_i) (1 - 2 \eta_j)}{2} \\
& \qquad  \quad + \left( \frac12 - m \right) ((1 - 2 \eta_i) + \delta_\kappa (1 - 2 \eta_j)) \, .
 \end{aligned}
\end{align}
Here $c_m^{\delta_\kappa}$ are coefficients of order $\mathcal{O}((c^{(g)})^0)$.
Since the expression is independent of $p$, we consider again the identity block with $\eta_p=0$, which are normalized by the two-point functions as in \eqref{Cij}. Note that the vacuum state is included in the Hilbert space of analytically continued minimal model.
Performing the monodromy move of $z$ from 0 to 1 and then from 1 to 0 only for the holomorphic part of $C_{ij}(z, \bar z)$, the absolute value of four-point function schematically becomes
\begin{align}
  \label{four2cb}
 & \left| \left \langle \mathcal{O}^\dagger_i(\infty) \mathcal{O}^\dagger_j (1) \mathcal{O}_i (z) \mathcal{O}_j (0) \right \rangle  \right| \\
  &\sim  \sum_{\delta_i , \delta_j ,\delta_\kappa = \pm } \sum_m \left| \left \langle \mathcal{O}^\dagger_i\mathcal{O}_i \right \rangle_{\delta_i} \left \langle
  \mathcal{O}^\dagger_j\mathcal{O}_j \right \rangle_{\delta_j} \exp \left( \frac{\pi c^{(g)} \kappa_{m}^{\delta_\kappa}}{3}  \right) \right| \, . \nonumber
 \end{align}
 The two-point function is written as the sum of $\left \langle
  \mathcal{O}^\dagger_j\mathcal{O}_j \right \rangle_{\delta_j}$ $(\delta_j = \pm)$, where
 \begin{align}
 \left| \left \langle
  \mathcal{O}^\dagger_j\mathcal{O}_j \right \rangle_{\delta_j} \right|
  \sim \exp\left( \delta_j  \frac{\pi c^{(g)}}{6} (1 - 2 \eta_j)\right) \, .
  \label{kappadelta}
\end{align} 
The saddles of conformal blocks are given as in \eqref{cbsaddle}, and hence $m$ should depend on $\delta_i,\delta_j$ labeling the saddles of two-point functions.
We may choose $m = (\delta_i +1)/2 $ and $m = (\delta_j \delta_\kappa +1)/2 $ such that the terms linear in $(1 - 2 \eta_i)$ and $(1 - 2 \eta_j)$ come from the two-point functions in the right hand side of \eqref{four2cb}.
We then reproduce \eqref{Sij} with the modular $S$-matrix of Liouville theory among non-degenerate operators.

\section{Higher-spin generalization}

Replacing the gauge group SL$(2,\mathbb{C})$ with SL$(N,\mathbb{C})$ in Chern-Simons action, the previous analysis can be extended to a higher-spin gravity, whose classical behavior can be captured by $\mathfrak{sl}(N)$ Toda theory with large central charge \cite{Hikida:2021ese,Hikida:2022ltr,Chen:2022ozy,Chen:2022xse}. 
For this extension, we adopt the following notations of $\mathfrak{sl}(N)$ Lie algebra. Let us denote the basis of $\mathbb{R}^N$ by $\epsilon_j$ $(j=1,2,\ldots,N)$ satisfying $(\epsilon_i, \epsilon_j ) = \delta_{i,j}$. Then, the simple roots are given by 
$e_j = \epsilon_j - \epsilon_{j+1}$ $(j=1,2,\ldots,N-1)$, which satisfy $(e_i ,e_j) = K_{ij}$ with $K_{ij}$ being the Cartan matrix of $\mathfrak{sl}(N)$.
The fundamental weights $\omega_j$ $(j=1,2,\ldots,N-1)$ satisfy $(\omega_i , e_j) = \delta_{i,j}$ and are given by
$
\omega_j = \sum_{l=1}^j \epsilon_l - \frac{j}{N} \sum_{l=1}^N \epsilon_l 
$.
The Weyl vector $\rho$ is the half of the sum over all positive roots or equivalently the sum over fundamental weights as
$
 \rho = \sum_{j=1}^{N-1} \omega_j = \sum_{j=1}^N \rho_j  \epsilon_j 
$ with $\rho_j =  \frac{N+1}{2} -j $.

We first study the possible saddle points of $\text{SL}(N,\mathbb{C})$ Chern-Simons gauge theory.
As in the case with $N=2$, we classify the non-trivial saddles of Chern-Simons theory by the holonomy matrix \eqref{dShol}, see \cite{Gutperle:2011kf,Ammon:2012wc}.
For non-singular geometry, we require that the eigenvalues of $\Omega$ introduced in \eqref{dShol}
are $2 \pi i(\lambda_1 , \lambda_2 , \ldots , \lambda_N)$ with 
$
 \lambda_j = m_j + \rho_j 
$.
With even $N$ (odd $N$), $m_j \in \mathbb{Z}$ or $\mathbb{Z} + 1/2$ $(m_j \in \mathbb{Z})$ for all $j$ but with $\sum_j m_j = 0$.
Defining $[e_{ij}]_{kl} = \delta_{i,k} \delta_{j,l}$, the corresponding gauge configuration may be given in a diagonal form as
\begin{align}
\begin{aligned}
& a =  i \sum_{j=1}^N e_{jj} ( (\rho - \eta)_j d \phi +  \lambda_j d t_E) \, , \\
&\tilde  a = -  i \sum_{j=1}^N e_{jj} ( (\rho - \eta)_j d \phi - \lambda_j d t_E) \, .
\end{aligned}
\end{align}
Here $\eta = (\eta_1, \dots , \eta_N)$ with $\eta_j$ related to higher-spin charges of corresponding dS$_3$ black hole.
We set $0 \leq  \eta_j \leq \rho_j$ for $j=1,\ldots, \lfloor \frac{N+1}{2} \rfloor$ and $0 \geq  \eta_j \geq \rho_j$ for $\lfloor \frac{N+1}{2} \rfloor + 1 , \ldots , N$.
Defining
$
 \lambda= \sum_j \lambda_j \epsilon_j  
$, 
the Gibbons-Hawking entropy corresponding to the configuration can be evaluated as
\begin{align} \label{higherGH}
  S_\text{GH}^{(\lambda)} = \frac{\pi}{3} c^{(g)} \frac{(\rho - \eta ,\lambda)}{(\rho , \rho)} \, ,
\end{align}
see  \cite{Hikida:2022ltr} for the details.

We then move to the Toda theory and find out the set of saddle points. The Toda theory has a parameter $b$ as in the Liouville theory, and the central charge is given by $c \, (\equiv i c^{(g)}) = N - 1 + 12 (Q,Q)$ with $Q = (b + b^{-1})\rho$. We are interested in the large $c^{(g)}$ regime, which can be realized by small $b$ with $\text{Re}\, b^{-2} < 0$ as
\begin{align}
b^{-2} = \frac{i c^{(g)}}{N(N^2-1)} - \frac{1}{N(N+1)} + \mathcal{O} ((c^{(g)})^{-1}) \, .
\end{align}
In order to evaluate two-point functions, we adopt the result \eqref{S0j} to make an explanation short.
We can analyze in a way analogous to the Liouville case, which leads to the same conclusion as we will show in  \cite{long}.
Denoting $\alpha =  (\rho - \eta)/b$, the modular $S$-matrix of the Toda theory with large $c^{(g)}$ is given by \cite{Drukker:2010jp,Hikida:2022ltr}
\begin{align}
S_{0\alpha} \sim    \sum_{w \in W} \epsilon (w) \exp \left( \frac{\pi}{6} c^{(g)} \frac{( \rho -\eta,w(\rho)) }{(\rho , \rho)}  \right )    \, , 
\end{align}
where $W$ denotes the Weyl group of SU$(N)$ and $\epsilon (w)$ is a sign related to $w$. We can see that the possible saddles of Chern-Simons theory are with $\lambda = w(\rho)$.
The leading contribution is given by $ \lambda = \rho$ and the corresponding Gibbons-Hawking entropy is \eqref{higherGH} with $\lambda = \rho$.
Since higher-spin charges are known to be invariant under the action of Weyl group, see, e.g. \cite{Bilal:1991eu,Bouwknegt:1992wg}, all the sub-leading saddles has the same higher-spin charges. This means that the all solutions should have the same geometric interpretation as the leading one. We can check that the analysis reduces to the previous one for $N=2$.

\section{Discussion}

We examined dS$_3$ gravity described by SL$(2,\mathbb{C})$ Chern-Simons theory and its higher-spin generalization. In general, there can be too many complex saddles of gravity path integral and we determined the set to take by applying recently proposed holography \cite{Hikida:2021ese,Hikida:2022ltr,Chen:2022ozy,Chen:2022xse}. 
In this letter, we investigated geometry dual to Liouville two- and four-point functions.
It is an important future problem to systematically formulate how to describe generic complex geometry from dual CFT multi-point functions.

We further extended the result to higher-spin gravity described by SL$(N,\mathbb{C})$ Chern-Simons theory. We presented only partial result on Toda two-point functions here but we are planing to report on more detailed analysis in \cite{long}. In particular, we examine effects of higher-spin charges in dS$_3$ black hole (or cosmological background) along the line of \cite{Gutperle:2011kf,Ammon:2012wc}, see, e.g. \cite{Krishnan:2013zya} for a previous attempt. As was done in \cite{Hikida:2021ese,Hikida:2022ltr,Doi:2022iyj,Narayan:2022afv,Narayan:2015vda,Sato:2015tta,Doi:2023zaf}, quantum information quantities are useful to examine the properties of dS higher-spin gravity and its holography. We also would like to comment on them. 

In the introduction, we have introduced bulk fields $\chi_j$ and boundary fields $\phi_j$. In generic holography, they are not directly related, since the bulk fields $\chi_j$ are dual to the boundary operators $\mathcal{O}_j$ and not the boundary fields $\phi_j$. However, in the current situation, bulk fields are given by Chern-Simons gauge fields $A,\bar A$, and after taking the diagonal gauge, we may relate the diagonal components of $A , \bar A$ to the Liouville/Toda fields $\phi_j^L (z),\phi_j^R(\bar z)$, where $\phi_j (z, \bar z)= \phi_j^L (z) + \phi_j^R(\bar z)$. See \cite{Campoleoni:2017xyl} in the case of AdS$_3$. The precise map between the bulk and boundary degrees of freedom should be useful to make the geometrical interpretation of dual CFT much clearer.

For our analysis, we utilized the known exact answers of Liouville/Toda field theory in order to determine the allowable saddles of gravity theory. However, it is quite rare that exact answers are available for the CFT dual to gravity theory. Even so, as mentioned above, CFT is usual much well-formulated than quantum gravity, so our holographic method should work more generically. For instance, conformal bootstrap technique is largely developed these days (see \cite{Simmons-Duffin:2016gjk,Poland:2018epd} for reviews), and the technique could be useful for our purpose. In any cases, it is important problem to extend the current analysis to other complex gravity theories, like a higher-dimensional one in \cite{Anninos:2011ui}. 

Furthermore, Liouville/Toda correlators used in this letter only tell us the possible saddles of corresponding gravity solutions, and they do not say anything about the properties of other gravitational saddles. However, as explained in \cite{Harlow:2011ny}, the other saddles of Liouville field theory can be selected if the region of complex parameter $b$ is changed. We expect that some information on the other saddles can be obtained by carefully treating the expanding parameter, and we are currently working on a related topic. It is also an important future problem to generalize it such as to be applicable to other complex gravity theories.

\begin{acknowledgments}
We are grateful to Katsushi Ito, Tatsuma Nishioka, Shigeki Sugimoto, and Tadashi Takayanagi for useful discussions.  The work is partially supported by Grant-in-Aid for Transformative Research Areas (A) ``Extreme Universe'' No.\,21H05187. The work of H.\,Y.\,C. is supported in part by Ministry of Science and Technology (MOST) through the grant 110-2112-M-002-006-. The work of Y.\,H. is supported by JSPS Grant-in-Aid for Scientific Research (B) No.\,19H01896 and Grant-in-Aid for Scientific Research (A) No.\,21H04469. Y.\,T. is supported by Grant-in-Aid for JSPS Fellows No.\,22J21950. The work of T.\,U. is supported by JSPS Grant-in-Aid for Early-Career Scientists No.\,22K14042.
\end{acknowledgments}


%

\end{document}